\documentclass[aps, prd, twocolumn, amsmath, floats,floatfix, 10pt, superscriptaddress, nofootinbib,
showkeys]{revtex4}

\DeclareRobustCommand{\VAN}[3]{#2}
\let\VANthebibliography\thebibliography
\def\thebibliography{\DeclareRobustCommand{\VAN}[3]{##3}\VANthebibliography}

\usepackage{amssymb}
\usepackage{amsmath}
\usepackage{verbatim}
\usepackage{mathrsfs}
\usepackage{amsfonts}
\usepackage{latexsym}
\usepackage{epsfig}
\usepackage{color}
\usepackage{graphicx,subfigure}
\usepackage{units}
\usepackage{envmath}
\usepackage{natbib}
\usepackage{ctable}
\usepackage[T1]{fontenc}
\usepackage[colorlinks=true, linkcolor=red, citecolor=blue, urlcolor=black]{hyperref}

\begin{document}


\definecolor{orange}{rgb}{0.9,0.45,0}

\def\CovDev{D}
\def\Res{{\mathcal R}}
\def\Gammaflat{\hat \Gamma}
\def\metricflat{\hat \gamma}
\def\Dflat{\hat {\mathcal D}}
\def\part_n{\partial_\perp}

\def\Lie{\mathcal{L}}
\def\A{\mathcal{X}}
\def\Aphi{\A_{\phi}}
\def\hAphi{\hat{\A}_{\phi}}
\def\E{\mathcal{E}}
\def\Ham{\mathcal{H}}
\def\M{\mathcal{M}}
\def\R{\mathcal{R}}
\def\p{\partial}

\def\hg{\hat{\gamma}}
\def\hA{\hat{A}}
\def\hD{\hat{D}}
\def\hE{\hat{E}}
\def\hR{\hat{R}}
\def\hcA{\hat{\mathcal{A}}}
\def\hDelt{\hat{\triangle}}

\def\na{\nabla}
\def\dif{{\rm{d}}}
\def\non{\nonumber}
\newcommand{\erf}{\textrm{erf}}

\renewcommand{\t}{\times}

\newcommand{\EDV}[1]{\textcolor{magenta}{[{\bf EDV}: #1]}}

\newcommand{\MN}[1]{\textcolor{blue}{[{\bf MN}: #1]}}

\long\def\symbolfootnote[#1]#2{\begingroup%
\def\thefootnote{\fnsymbol{footnote}}\footnote[#1]{#2}\endgroup}


\title{Structure Formation in Various Dynamical Dark Energy Scenarios} 

\author{Masoume Reyhani}
\email{masoume.r1079@gmail.com}
\affiliation{Department of Physics, K.N. Toosi University of Technology, P.O. Box 15875-4416, Tehran, Iran}
\affiliation{PDAT Laboratory, Department of Physics, K.N. Toosi University of Technology, P.O. Box 15875-4416, Tehran, Iran}

\author{Mahdi Najafi}
\email{mahdinajafi12676@yahoo.com}
\affiliation{Department of Physics, K.N. Toosi University of Technology, P.O. Box 15875-4416, Tehran, Iran}
\affiliation{PDAT Laboratory, Department of Physics, K.N. Toosi University of Technology, P.O. Box 15875-4416, Tehran, Iran}

\author{Javad T. Firouzjaee}
\email{firouzjaee@kntu.ac.ir}
\affiliation{Department of Physics, K.N. Toosi University of Technology, P.O. Box 15875-4416, Tehran, Iran}
\affiliation{PDAT Laboratory, Department of Physics, K.N. Toosi University of Technology, P.O. Box 15875-4416, Tehran, Iran}
\affiliation{School of Physics, Institute for Research in Fundamental Sciences (IPM), P.O. Box 19395-5531, Tehran, Iran}

\author{Eleonora Di Valentino}
\email{e.divalentino@sheffield.ac.uk}
\affiliation{School of Mathematics and Statistics, University of Sheffield, Hounsfield Road, Sheffield S3 7RH, United Kingdom}

\today

\begin{abstract} 
    This research investigates the impact of the nature of Dark Energy (DE) on structure formation, focusing on the matter power spectrum and the Integrated Sachs-Wolfe effect (ISW). By analyzing the matter power spectrum at redshifts $z = 0$ and $z = 5$, as well as the ISW effect on the scale of $\ell = 10-100$, the study provides valuable insights into the influence of DE equations of state (EoS) on structure formation. The findings reveal that dynamical DE models exhibit a stronger matter power spectrum compared to constant DE models, with the JBP model demonstrating the highest amplitude and the CPL model the weakest. Additionally, the study delves into the ISW effect, highlighting the time evolution of the ISW source term $\mathcal{F}(a)$ and its derivative $d\mathcal{F}(a)/da$, and demonstrating that models with constant DE EoS exhibit a stronger amplitude of $\mathcal{F}(a)$ overall, while dynamical models such as CPL exhibit the highest amplitude among the dynamical models, whereas JBP has the lowest. The study also explores the ISW auto-correlation power spectrum and the ISW cross-correlation power spectrum, revealing that dynamical DE models dominate over those with constant DE EoS across various surveys. Moreover, it emphasizes the potential of studying the non-linear matter power spectrum and incorporating datasets from the small scales to further elucidate the dynamical nature of dark energy. This comprehensive analysis underscores the significance of both the matter power spectrum and the ISW signal in discerning the nature of dark energy, paving the way for future research to explore the matter power spectrum at higher redshifts and in the non-linear regime, providing deeper insights into the dynamical nature of dark energy.

\end{abstract}




\maketitle

\vspace{0.8cm}

\section{Introduction}\label{sec:1} 
It would not be an exaggeration to call the discovery of the expanding universe, the triumph of modern cosmology~\cite{riess1998observational,perlmutter1999measurements}. This expansion could be explained through the inclusion of Dark Energy (DE) component, which contributes to nearly 70\% of the energy budget in the standard $\Lambda$CDM model. This model posits the universe as homogeneous and isotropic~\cite{di2023cosmological, peebles1982large}. 
As the standard cosmological model, the Lambda cold dark matter ($\Lambda$CDM) model has proven highly successful in explicating a broad spectrum of cosmological observations. These include the cosmic microwave background (CMB)~\cite{nolta2004first, spergel2007three, ade2016planck, aghanim2020planck}, type Ia supernovae~\cite{riess1998observational, perlmutter1999measurements}, galaxy surveys~\cite{tegmark2006cosmological, eisenstein2005detection}, and weak lensing~\cite{hildebrandt2017kids, abbott2018dark}. 

However, in recent years, significant tensions have emerged in observational data, particularly concerning the Hubble parameter  $H_0$~\cite{di2021realm,krishnan2021running,guo2019can, abdalla2022cosmology, di2021snowmass2021, perivolaropoulos2022challenges} and the amplitude of late-time matter clustering $\sigma_8-S_8$~\cite{di2018reducing,kazantzidis2021sigma, di2021cosmology, di2021snowmass2021}. These tensions suggest that the $\Lambda$CDM model may not be the complete picture. The Hubble tension is a discrepancy between the direct and indirect measurements of Hubble constant $H_0$~\cite{bernui2023solution,aghanim2020planck,yang2023dynamics,aiola2020atacama,dutcher2021measurements,verde2019tensions,di2021realm,mcdonough2022early, madhavacheril2023atacama, balkenhol2023measurement, huang2020hubble,riess2022comprehensive,riess2022cluster,murakami2023leveraging}, and the $\sigma_8-S_8$ tension which represents the disagreement between the measurements of late-time and early-time amount of matter clustering~\cite{perivolaropoulos2022challenges, asgari2021kids,mcdonough2022early,abdalla2022cosmology,li2016measurement,gil2016clustering,aiola2020atacama,hu2023hubble}. 
Moreover, anomalies in CMB anisotropy~\cite{rassat2014planck, akrami2020planck,schwarz2016cmb, perivolaropoulos2014large, cayuso2020towards} and CMB cold spots~\cite{vielva2004detection, vielva2004detection, cruz2007non, perivolaropoulos2022challenges}
have spurred discussions among the scientific community~\cite{hu2023hubble, cayuso2020towards}. Numerous studies have been conducted to address these discrepancies~\cite{di2021realm}, including models that challenge fundamental properties of dark matter(DM) or dark energy (DE). These investigations encompass aspects such as the mass-temperature relation~\cite{del2019mass,naseri2020effect} and scaling relation~\cite{blanchard2021closing,naseri2021super} of galaxy clusters.

To address these tensions, one approach involves investigating potential systematic errors. The consistency between independent direct and indirect measurements, combined with the enhanced precision of measurement instruments, underscores the need to explore alternative solutions to resolve these cosmological crises. It is increasingly apparent that finding a resolution may necessitate delving into new physics paradigms to accurately describe the universe~\cite{di2023cosmological, natoli2018exploring}.
One of the most promising ways to address these tensions is to consider dynamical DE models~\cite{zhao2017dynamical,pan2019interacting,yang2019observational}, which allow the equation of state of DE to vary with time. Dynamical DE models can be motivated by a variety of theoretical considerations, such as the need to explain the late-time acceleration of the Universe, the existence of a DE field that couples to dark matter or other fields, or the possibility that DE is not a fundamental constant but rather a dynamical quantity that evolves with time~\cite{yang2021dynamical, chevallier2001accelerating, linder2003exploring,efstathiou1999constraining, jassal2005observational,barboza2008parametric}. Some of the other alternative DE models are Early DE~\cite{gomez2022coupled, escudero2022early, karwal2022chameleon, herold2022new, jiang2022toward, smith2022hints, reeves2023restoring,Efstathiou:2023fbn,Kamionkowski:2022pkx,Poulin:2023lkg,Niedermann:2023ssr,Eskilt:2023nxm,Gsponer:2023wpm,Goldstein:2023gnw}, Ginzburg-Landau Theory of DE~\cite{banihashemi2019ginzburg, banihashemi2022fluctuations} and interacting Dark Energy-Dark Matter (IDMDE)~\cite{naseri2020effect, farrar2004interacting,Valiviita:2008iv,Gavela:2009cy,Salvatelli:2014zta,DiValentino:2017iww,Kumar:2017dnp,
Martinelli:2019dau,Yang:2019uog,DiValentino:2019ffd,Pan:2019jqh,Kumar:2019wfs,Kumar:2016zpg,Murgia:2016ccp,Pourtsidou:2016ico,Lucca:2020zjb,Gomez-Valent:2020mqn,Yang:2020uga,Nunes:2021zzi,Nunes:2022bhn,Zhai:2023yny,Escamilla:2023shf}.

An intriguing avenue for constraining DE models is to study the Integrated Sachs-Wolfe (ISW) effect within observational data. This entails determining the cross-correlation between the ISW signal and the distribution of galaxies. The ISW effect is a cosmological phenomenon that occurs for the time dependence of the gravitational potential as photons propagate through the observable Universe. It is caused by the redshift or blueshift of photons as they pass through regions of space where the gravitational potential undergoes changes. In this respect, calculation of the ISW effect can be a convenient measure for studying DE and can probe some modifications of the $\Lambda$CDM model such as interacting dark sector models and modified gravity~\cite{kovacs2019more,yengejeh2023integrated,krolewski2022integrated,ghodsi2022integrated,kable2022probing,Seraille:2024beb}. The ISW effect is especially sensitive to the time-varying nature of the DE equation of state, making it a unique probe of dynamical DE models. 

In this paper, we study the effect of the nature of DE on structure formation across a spectrum of DE models. We start with a review of the various DE models and, subsequently, we conduct a comparative assessment of the introduced models against the standard $\Lambda$CDM model. To gain insights into the nature of Dark Energy, we adopt two distinct approaches. 
Firstly, our exploration involves an examination of the linear matter power spectrum. This analysis reveals that scrutinizing this spectrum is instrumental in distinguishing between various DE models. In a second step, our focus shifts to the calculation of the ISW signal amplitude within each model, compared against the predictions of the standard $\Lambda$CDM model.  
For this purpose, it has been suggested~\cite{schafer2008integrated,yengejeh2023integrated,ghodsi2022integrated} that ISW signal auto-correlation and cross-correlation with the galaxy distribution is a powerful tool to differentiate proposed models from one another.

Our results suggest the following: 1. Each DE model hints at the distinct structure formation pattern in the matter power spectrum. 2. The ISW effect could be a powerful tool for constraining DE models and distinguishing them from the $\Lambda$CDM model. Future observations of the ISW effect, especially when combined with other cosmological probes, could provide valuable insights into the nature of DE and its evolution over time.

The outline of this work is presented as follows. In Sec.~\ref{sec:2}, we illustrate the theoretical framework and the DE models under the study. In Sec.~\ref{sec:3}, we briefly describe the methodology of the observational analysis tools and the likelihoods of this study. Then in Sec.~\ref{sec:4}, we study the matter power spectrum for these models. Furthermore, in Sec.~\ref{sec:5}, we calculate the ISW effect within the context of these DE models and compare our results with that of the $\Lambda$CDM model. Finally, we scrutinize the results and summarize the findings in Sec.~\ref{sec: 6}.

\section{Dynamical Dark Energy}\label{sec:2}

At very large scales ($\sim 100$ Mpc) the universe can be assumed homogeneous and isotropic. By considering a flat expanding Friedmann-Lemaître-Robertson-Walker (FLRW), the line element is described by 
\begin{equation}
ds^2=a^2(\tau)[-d\tau^2+d{ \bf x}^2]
\end{equation}
where $a(\tau)$ is the expansion factor of the universe known as scale factor, and $\tau$ is the conformal time defined as $d\tau=dt/da$. For this universe the Einstein equations are given by~\cite{ma1995cosmological}: 
\begin{equation} \label{eq:1}
\mathcal{H}^2(a)\equiv(\frac{{a^\prime}}{a})^2=\frac{8\pi G}{3}a^2\rho_{\rm i}
\end{equation}
\begin{equation} \label{eq:2}
\frac{d}{d\tau}(\frac{a^\prime}{a})=\frac{-4\pi G}{3}a^2(\rho_{\rm i}+3P_{\rm i})
\end{equation}
where the primes denote the derivation with respect to conformal time $\tau$, $\mathcal{H}(a) =a^\prime/a$ is defined as the conformal Hubble parameter which represents the expansion rate of the universe, $G$ is the Newtonian constant, and $\rho_{\rm i}$ is the energy density of non-interacting components of the universe (i.e. radiation represented by $\rho_{\rm r}$, DE represented by $\rho_{\rm DE}$, dark matter represented by $\rho_{\rm c}$, and baryonic matter represented by $\rho_{\rm b}$). The same notation applied to $ P_{\rm i}$ as well, which describes the pressure of each aforementioned component. Also, the relation between density and pressure is described by $w_{\rm i}=P_{\rm i}/\rho_{\rm i}$ which is called barotropic equation of state. For the components studied in the paper we have, $w_{\rm r}=P_{\rm r}/\rho_{\rm r}=1/3$, $w_{\rm c}=P_{\rm c}/\rho_{\rm c}=0$, $w_{\rm b}=P_{\rm b}/\rho_{\rm b}=0$, $w_{\rm DE}=P_{\rm DE}/\rho_{\rm DE}=-1$. The cosmic fluid can be presented as a perfect fluid with the following energy momentum :
\begin{equation} \label{eq:3}
T^\mu_\nu=Pg^\mu_\nu+(\rho+P)U^\mu U_\nu
\end{equation}
Considering the FLRW metric, Eq.~(\ref{eq:3}) leads to $T^\mu_\nu=diag(-\rho,P,P,P)$. From Bianchi identity, $\nabla_\mu G^\mu_{\nu}=0$ is obtained, which implies the conservation of energy-momentum:
\begin{equation} \label{eq:4}
\nabla_\mu T_{\nu} ^\mu=0
\end{equation}
Considering $\nu=0$, the Eq.~(\ref{eq:4}) will return the continuity equation which simply depicts the time evolution of the components of the universe: 
\begin{equation} \label{eq:5}
\rho^\prime+3\mathcal{H}(\rho+P)=0,
\end{equation}
\\or
\begin{equation} \label{eq:6}
\rho^\prime+3\mathcal{H}(1+w)\rho = 0.
\end{equation}
We have ignored the subscript $i$, since this equation is true for every component of the non-interacting universe.
\\With all that knowledge, the Eq.~(\ref{eq:1}) can be written as below~\cite{yang2021dynamical}:
\begin{eqnarray} \label{eq:7}
\mathcal{H}^2(a)=\frac{8\pi G}{3}a^2[\rho_{\rm r,0} a^{-4}+\rho_{\rm b,0} a^{-3}+\rho_{\rm c,0} a^{-3}\nonumber\\
+\rho_{\rm DE,0} (\frac{a}{a_0})^{-3} exp (-3 \int_{\rm a_0}^{a}\frac{w_{\rm DE}(a')}{a'}da')]
\end{eqnarray}
Where subscript $0$ indicates the current value of each component. Considering $w_{\rm DE} = -1$, Eq.~(\ref{eq:7}) simply returns the standard $\Lambda$CDM model. 

By considering different scenarios for DE equation of state, we aim to investigate the impact of the nature of DE equation of state on the matter power spectra, the evolution of gravitational potential, and subsequently the ISW effect. To this end, we will describe the DE models under study. A key illustrative tool throughout this work will be the comparison of the introduced models to the standard $\Lambda$CDM model and among themselves.
\begin{itemize}
\item \textbf{Model 1:} In this model, we consider the DE equation of state to be constant and $w_{\rm DE} = w_0 \neq -1$. We will call this model $w$CDM.
\item \textbf{Model 2:} The first dynamical DE model under study is the well-known Chevallier-Polarski-Linder parametrization, henceforth referred to as CPL~\cite{chevallier2001accelerating, linder2003exploring}:
\begin{equation} \label{eq:8}
w_{\rm DE}(a)=w_0 +w_{\rm a}(1-a)
\end{equation}
\item \textbf{Model 3:} The second dynamical DE equation of state is proposed by Jassal-Bagla-Padmanabhan and is referred to as JBP~\cite{jassal2005observational} in this paper:
\begin{equation} \label{eq:9}
w_{\rm DE}(a)=w_0 + w_{\rm a} a (1-a)
\end{equation}
\item \textbf{Model 4:} The last model being under scrutiny in this article is proposed by Barboza-Alcaniz, and it is known as BA~\cite{barboza2008parametric}:
\begin{equation} \label{eq:10}
w_{\rm DE}(a)=w_0 +w_{\rm a}(\frac{1-a}{2a^2-2a+1})
\end{equation}
\end{itemize}

In order to study the time evolution of the potential, it is necessary to consider perturbations to the FLRW metric. In synchronous gauge we can reconsider the line element to be as~\cite{ma1995cosmological}: 
\begin{equation} \label{eq:11}
ds^2=a^2(\tau)[-d\tau^2+(\delta_{\rm ij}+h_{\rm ij})dx^idx^j]
\end{equation}
The metric perturbation $h_{\rm ij}$ can be decomposed into a trace part $h\equiv h_{\rm ii}$ and a traceless part consisting of three pieces: $h_{\rm ij}^\parallel$, $h_{\rm ij}^\perp$ and $h_{\rm ij}^T$. By moving to Fourier space, the equations governing the dimensionless density perturbations $\delta_{\rm i}=\delta\rho_{\rm i}/\rho_{\rm i}$ and velocity perturbations $\theta_{\rm i}=\partial_jv_i^j$ are given by: 
\begin{eqnarray} \label{eq:12}
\delta^\prime_{\rm DE} =-(1+w_{\rm DE})(\theta_{\rm DE}+\frac{h^\prime}{2})-3\mathcal{H}(c^2_{\rm sDE}-w_{\rm DE})\nonumber\\
\times[\delta_{\rm DE}+3\mathcal{H}(1+w_{\rm DE})\frac{\theta_{\rm DE}}{k^2}]-3\mathcal{H}w^{\prime}_{\rm DE}\frac{\theta_{\rm DE}}{k^2}
\end{eqnarray}
\begin{equation} \label{eq:13}
\theta^\prime_{\rm DE}=-\mathcal{H}(1-3c^2_{\rm sDE})\theta_{\rm DE}+\frac{c^2_{\rm sDE}}{(1+w_{\rm DE})}k^2\delta_{\rm DE}
\end{equation}
\begin{equation} \label{eq:14}
\delta^\prime_{\rm c}=-(\theta_{\rm c}+\frac{h^\prime}{2})
\end{equation}
\begin{equation} \label{eq:15}
\theta^\prime_{\rm c}=-\mathcal{H}\theta_{\rm c}
\end{equation}
Where the $c^2_{\rm sDE}$ is the squared sound speed of DE component in rest frame, $c^2_{\rm sDE}=\delta P_{\rm DE}/\delta \rho_{\rm DE}$, and $k$ is the wavenumber in Fourier space. It is also worth to define adiabatic sound speed, $c^2_{\rm aDE}=dP_{\rm DE}/d\rho_{\rm DE}=w_{\rm DE}+ w^\prime_{\rm DE}/(\rho^\prime_{\rm DE}/\rho_{\rm DE})$. Considering a barotropic equation of state will lead to the $c^2_{\rm aDE}=c^2_{\rm sDE}=w_{\rm DE}$. By considering DE as an adiabatic fluid, we will have $c^2_{\rm aDE}=c^2_{\rm sDE}=w_{\rm DE}<0$ which as a consequence will lead to instabilities in DE fluid. In order to solve this problem we will consider $c^2_{\rm sDE}=1$.

Having outlined the framework within which we will operate, it would be illustrative to delineate the methodology that will assist us in constraining the cosmological parameters.

\section{Methodology}\label{sec:3}
In the following, we briefly describe the observational analysis tools and likelihoods used in this study. Using a modified version of the publicly available cosmological code \texttt{CAMB}~\cite{lewis2000efficient, howlett2012cmb}, and by means of the publicly available Monte-Carlo Markov Chain \texttt{Cobaya}, we successfully obtained the best fit to the data~\cite{powell2009bobyqa,cartis2018escaping,cartis2019improving,torrado2021cobaya}.\\ For the base cosmological model, $\Lambda$CDM, we have used the following 6 parameter space:\\
\begin{equation}\label{eq:16}
\mathcal{P}=\{\Omega_{\rm b}h^2,\Omega_{\rm c}h^2,100\theta_{\rm MC},\tau,n_{\rm s},\ln[10^{10}A_{\rm s}]\}
\end{equation} 
For $w$CDM model, we have added an extra degree of freedom in the form of $w_0$:
\begin{equation}\label{eq:17}
\mathcal{P}=\{\Omega_{\rm b}h^2,\Omega_{\rm c}h^2,100\theta_{\rm MC},\tau,n_{\rm s},\ln[10^{10}A_{\rm s}],w_0\}
\end{equation} 
And for the dynamical parameterization [Eqs. (\ref{eq:8})-(\ref{eq:10})], we have added two extra degrees of freedom, $w_{\rm a}$ and $w_0$ to the standard, 6 parameter space:
\begin{equation}\label{eq:18}
\mathcal{P}=\{\Omega_{\rm b}h^2,\Omega_{\rm c}h^2,100\theta_{\rm MC},\tau,n_{\rm s},\ln[10^{10}A_{\rm s}],w_0,w_{\rm a}\}
\end{equation}
The baseline likelihoods we used include~\cite{aghanim2020plancklikelihoods,aghanim2020plancklensing}:
\begin{itemize}
\item \texttt{\textbf{Commander}} likelihood which provides low multipoles TT data in the range ($2\leq \ell \leq 29$).
\item \texttt{\textbf{SimAll}} likelihood which provides low multipoles EE data in the range ($2\leq \ell \leq 29$).
\item \texttt{{\tt Plik}} TT,TE,EE likelihood which provides the high multipoles TT, TE, and EE data in the range ($30\leq \ell \lesssim 2500$ for TT and $30\leq \ell \lesssim 2000$ for TE and EE).
\item \textbf{lensing} reconstruction likelihood which is obtained with a trispectrum analysis and can provide complementary information to the Planck CMB power spectra.
\end{itemize}

\begin{figure}[t!]
\begin{minipage}{1\linewidth}
\includegraphics[width=\columnwidth]{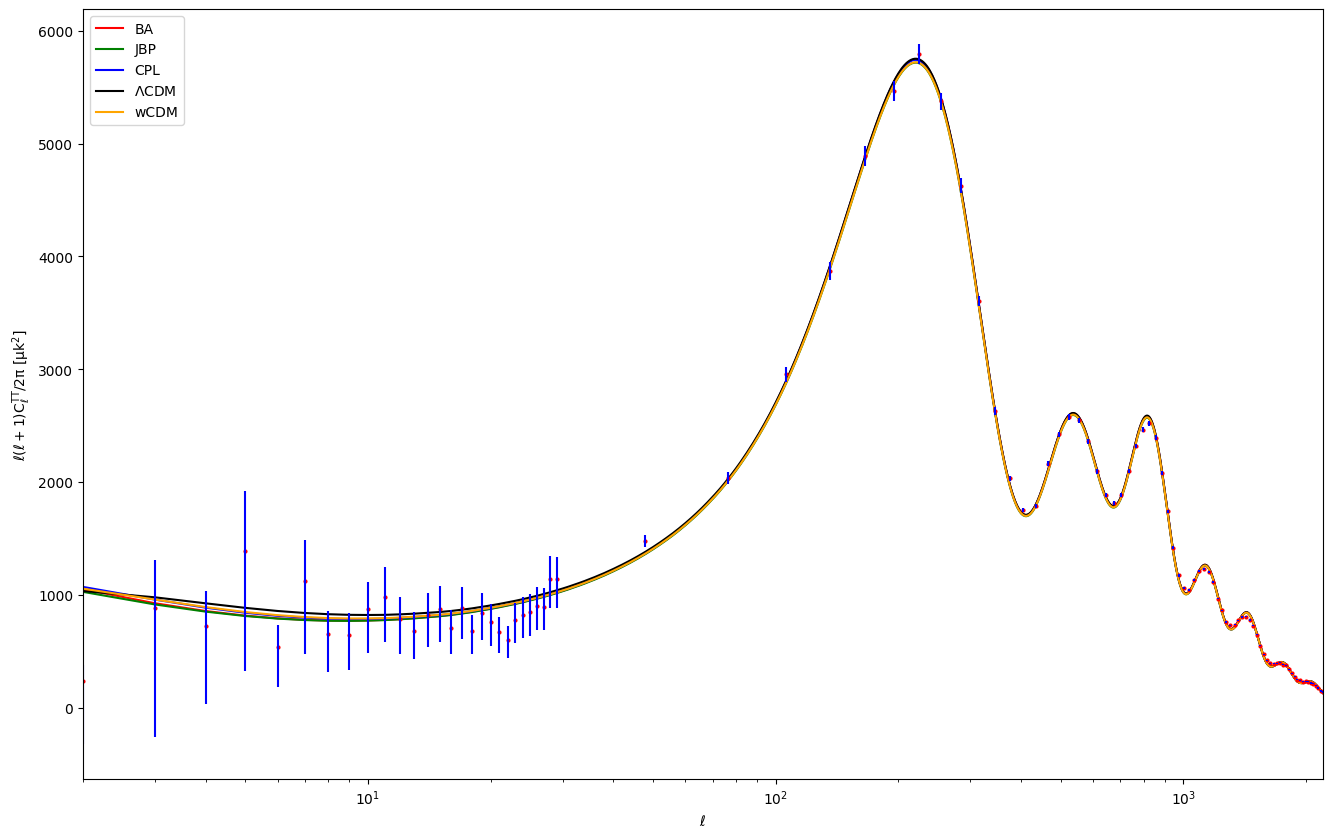} 
\caption{Comparison of the CMB temperature power spectrum of each model under the study (BA: Red, JBP: Green, CPL: Blue, $\Lambda$CDM: Black, $w$CDM: Yellow) based on the best-fit values obtained from the cosmological code \texttt{Cobaya} assuming the baseline likelihoods with Planck 2018 (blue data points with error bars).}
\label{fig:fig1}
\end{minipage}
\end{figure}

Table~\ref{table:priors} shows the flat priors used to obtain best-fit of the data. Table~\ref{tab:my_label} shows the best-fit values for parameters used in each DE parameterization. It has to be emphasized that in $w$CDM case, our best-fit analysis relies on Planck baseline likelihoods listed above, and the value for $w_0$ resides in the phantom region. We can see in Fig.~\ref{fig:fig1} the temperature power spectrum of each model obtained with the best-fit results, and it depicts the agreement of these best-fits for each DE parameterization used in this study.

By understanding the methodology and datasets, one would be able to study the DE equation of state effects on structure formation. To do so, we will start by studying the effect of the impact of DE equation of state on the matter power spectrum.

\begin{table}[!ht]
\begin{center}
\renewcommand{\arraystretch}{1.5}
\begin{tabular}{l@{\hspace{0. cm}}@{\hspace{2 cm}} c}
\hline\hline
\textbf{Parameter} & \textbf{Prior} \\
\hline\hline
$\Omega_{\rm b} h^2$ & $[0.005\,,\,0.1]$ \\
$\Omega_{\rm c} h^2$ & $[0.001\,,\,0.99]$ \\
$\tau$ & $[0.01, 0.8]$ \\
$100\,\theta_{\rm MC}$ & $[0.5\,,\,10]$ \\
$\log(10^{10}A_{\rm S})$ & $[1.61\,,\,3.91]$ \\
$n_{\rm s}$ & $[0.8\,,\, 1.2]$ \\
$w_0$ & $[-3\,,\,1]$ \\
$w_{\rm a}$ & $[-3\,,\,2]$\\
\hline\hline
\end{tabular}
\caption{Ranges for the flat prior distributions on cosmological parameters in the study. For $w$CDM, we used [-3\,,\,-0.333] to impose an expanding universe.}
\label{table:priors}
\end{center}
\end{table}

\begin{table*}[t] 
\caption{The best-fit dataset from our baseline dataset used to study each DE model.}
\centering 
\begin{tabular}{c||c|c|c|c|c}
\hline\hline
\textbf{Model} & \textbf{$\Lambda CDM$} &\textbf{$w CDM$} & \textbf{CPL} & \textbf{JBP} & \textbf{BA} \\
\hline\hline
\textbf{$\Omega_{\rm b}h^2$} & $0.022340$ & $0.02241$ & $0.02242$ & $0.02246$ & $0.02244$ \\
\textbf{$\Omega_{\rm c}h^2$} & $0.11984$ & $0.11902$ & $0.11863$ & $0.11858$ & $0.11860$\\
\textbf{$H_0$} & $67.36$ & $81.31$ & $82.42$ & $99.79$ & $96.51$ \\
\textbf{$10^9A_{\rm S}$} & $2.111$ & $2.103$ & $2.104$ & $2.095$ & $2.103$ \\
\textbf{$n_{\rm s}$} & $0.9636$ & $0.9663$ & $0.9662$ & $0.9673$ & $0.9677$ \\
\textbf{$\tau$} & $0.059$ & $0.058$ & $0.058$ & $0.058$ & $0.057$ \\
\textbf{100$\theta_{\rm MC}$} & $1.04092$ & $1.04101$ & $1.04104$ & $1.04106$ & $1.04105$ \\
\textbf{$w_0$} & $-1$ & $-1.42$ & $-0.92$ & $-1.85$ & $-1.31$ \\
\textbf{$w_{\rm a}$} & $-$ & $-$ & $-2.53$ & $-0.38$ & $-1.69$ \\
\hline\hline
\end{tabular}
\label{tab:my_label}
\end{table*}

\section{Matter power spectrum}\label{sec:4}
In order to illustrate the effect of dynamical DE on structure formation over time, we will study how the dynamical nature of DE affects the amplitude of the matter power spectrum. By analyzing the matter power spectrum, we can also discern the effect of DE across various scales. Here, we will focus on the linear evolution of the matter power spectrum in different redshifts, but studying nonlinear correction to the matter power spectrum could unveil interesting results.

\subsection{Linear matter power spectrum}\label{subsec:4.1}
To study the statistical correlations, the linear matter power spectrum is a powerful tool for this purpose~\cite{dodelson2020modern, orjuela2023machine}. The matter power spectrum for any given redshift can be written as a function of redshift $z$ and wavenumber $k$: 
\begin{equation} \label{eq:19}
P_l (k,z)= \frac{8\pi^2}{25}\frac{\mathcal{A}_{\rm s}}{\Omega^2_{\rm m}} D_+ ^2(z) T^2 (k) \frac{k^{n_{\rm s}}}{H_0^4 k_{\rm p}^{n_{\rm s} -1}}
\end{equation}
Where $\mathcal{A}_{\rm s}$ and $n_{\rm s}$ represent the primordial amplitude and spectral index respectively. $T(k)$ is the transfer function, where on large-scale $T(k)=1$. $k_{\rm p}$ is an arbitrary pivot scalar, and $H(a)=\Dot{a}/a$ is the Hubble parameter, where the dot denotes the derivative with respect to physical time $t$. As before, $H_0$ denotes the present value of the Hubble parameter, known as the Hubble constant.

To scrutinize the evolution of structure formation, it is descriptive to consider the matter power spectrum at different redshifts.

\begin{figure}[t!]
\begin{minipage}{1\linewidth}
\includegraphics[width=\columnwidth]{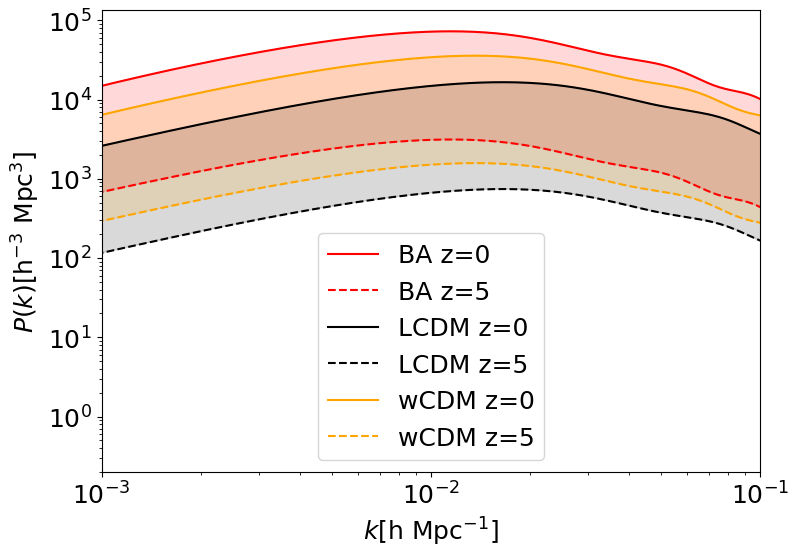} \\~\\
\includegraphics[width=\columnwidth]{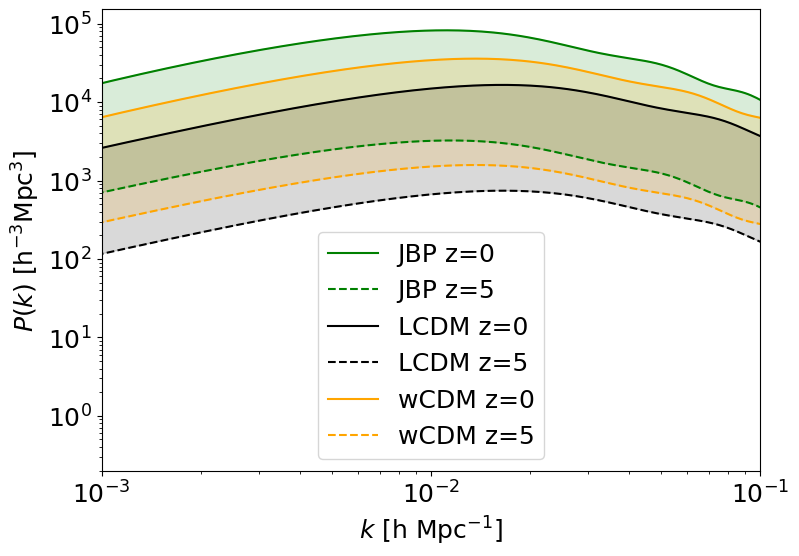} \\~\\
\includegraphics[width=\columnwidth]{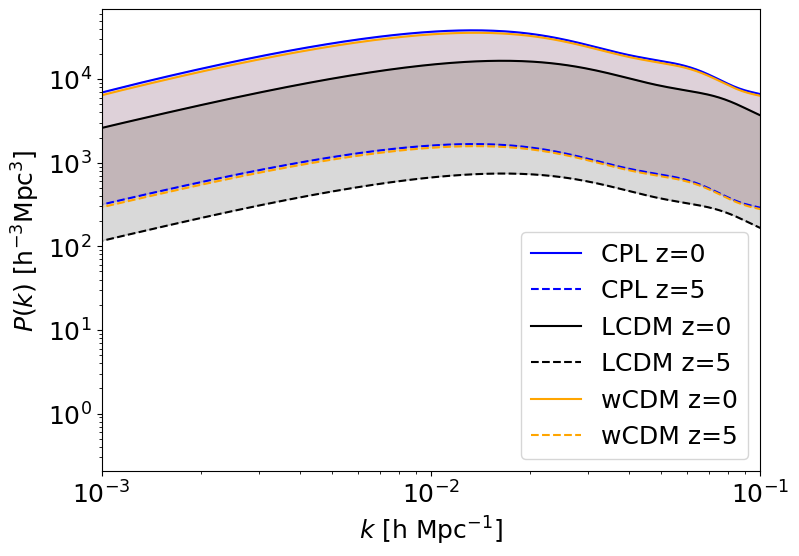} 
\caption{The matter power spectrum as a function of comoving wavenumber $k$ for 3 dynamical DE equation of state models (Top: BA, Middle: JBP, Bottom: CPL), compared with 2 constant DE equation of state model ($\Lambda$CDM and $w$CDM), at redshift $z=0$ (solid lines) and redshift $z=5$ (dashed lines). The shaded area in each panel represents the range within which the matter power spectrum of redshifts between $0<z<5$ is contained.}
\label{fig:fig2}
\end{minipage}
\end{figure}

In Fig.~\ref{fig:fig2}, we present a comparison of the matter power spectrum as a function of comoving wavenumber $k$ among dynamical dark energy (DE) equation of state parameterizations --i.e., the BA, JBP, and CPL models -- and the constant dark energy equation of state model $w$CDM, along with the standard $\Lambda$CDM model. This comparison is based on the best fit of these scenarios to our baseline likelihoods. 
All three panels demonstrate that regardless of the chosen model, the general amplitude of the matter power spectrum is suppressed as we move towards higher redshifts (dotted lines). This observation arises because structure formation is stronger at lower redshifts, closer to the current age of the universe. It can also be concluded that in all three cases of dynamical dark energy, regardless of redshift, the general amplitude of the matter power spectrum has higher values compared to those with a constant dark energy equation of state. One possible explanation for this behavior lies in the dynamical nature of dark energy; considering dark energy to be dynamical leads to a stronger effect from the matter component in the evolution history of the universe.
As indicated in Fig.~\ref{fig:fig2}, $w$CDM has a stronger general amplitude compared to the $\Lambda$CDM model for the same redshifts. In the top panel of Fig.~\ref{fig:fig2}, one can also observe a slight shift in the Baryon Wiggles peaks towards lower wavenumbers $k$s~(larger scales) in the BA model compared to constant DE EoS scenarios. The middle panel of Fig.~\ref{fig:fig2} shows the same behaviour for the JBP parameterisation. However, the bottom panel of Fig.~\ref{fig:fig2} shows an interesting attribute. The matter power spectrum of the CPL model, though slightly higher in values in general, behaves similarly to the $w$CDM model, i.e. the shift in Baryon Wiggles peaks is not as noticeable as in two other cases. Also, it can be seen that distinguishing between the CPL and $w$CDM models is challenging.
Furthermore, one can observe a shift in $k_{eq}$ (the wavenumber representing the transition from a radiation-dominated to a matter-dominated universe) towards lower $k$ values for BA and JBP compared to $w$CDM and $\Lambda$CDM. Comparing $w$CDM with $\Lambda$CDM, a clear shift of $k_{eq}$ towards lower $k$ values is noticeable in the case of $w$CDM. An interesting point to note is that distinguishing the shift in $k_{eq}$ between CPL and $w$CDM remains challenging.

\begin{figure} [t!]
\begin{minipage}{1\linewidth}
\includegraphics[width=\columnwidth]{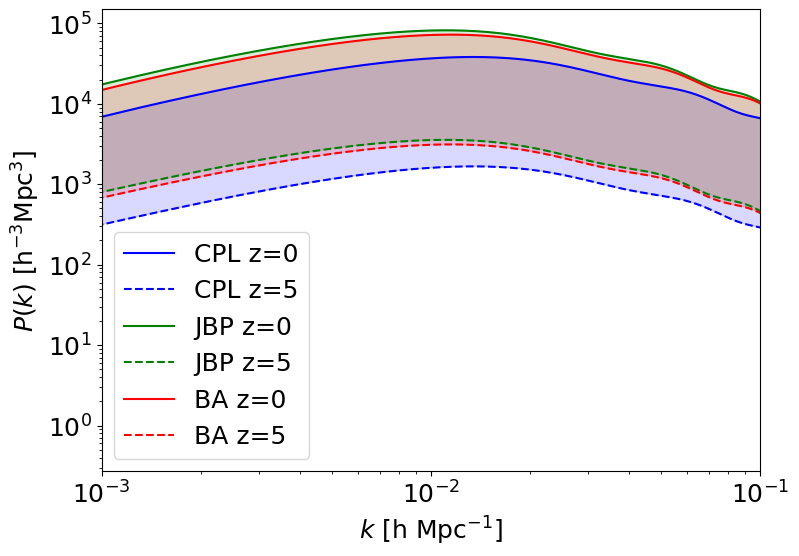} 
\caption{Comparison of the matter power spectra as a function of comoving wavenumber $k$ among 3 dynamical DE models in redshift $z=0$ (solid lines) and redshift $z=5$ (dashed lines). The shaded area represents the range within which the matter power spectrum for redshifts between $0<z<5$ is expected.  }
\label{fig:fig3}
\end{minipage}
\end{figure}

Fig.~\ref{fig:fig3} depicts the evolution of these three dynamical DE models at redshifts $z=0$ and $z=5$. As observed, when we consider DE behavior as described by the CPL model, the amplitude of the matter power spectrum is the lowest compared to the BA and JBP parameterizations at the mentioned redshifts. However, the JBP and BA parameterizations exhibit an interesting similarity in the given redshifts. As depicted, the JBP and BA models are nearly indistinguishable. It should be noted that the shaded area in each panel of Fig.~\ref{fig:fig2},~\ref{fig:fig3} represents the range within which the matter power spectrum for redshifts between $0<z<5$ is expected. 

It is crucial to note that considering different dataset combinations may yield diverse behaviors in the matter power spectrum. Particularly, incorporating datasets that observe redshifts deep into the matter-dominated era could provide valuable insights into the dynamical nature of the BA, JBP, and CPL models.

Gaining insight into the impact of the nature of DE on the matter power spectrum enables a more intuitive exploration of the Integrated Sachs-Wolfe (ISW) signal.

\section{Measuring Integrated Sachs-Wolfe effect}\label{sec:5}

Here, we will focus on two changes in the wavelength of Cosmic Microwave Background (CMB) radiation. The first is based on the fact that due to the inhomogeneity in the universe, the photons will encounter potential wells resulting in an increase in their energy or a decrease in their wavelength. The second change is due to the expansion of the universe which will cause photons to lose their energy. In other words, photons traveling toward us from the last scattering surface will experience a change in their wavelength to a larger value.
Based on what we have discussed, when a photon enters a potential well, it will experience a blueshift in its wavelength. However, due to the expansion of the universe, we will observe only a slight difference between the original and final wavelength of the photon.
\\ To calculate the ISW effect for each model, it is necessary to obtain the matter power spectrum (see Subsec.~\ref{subsec:4.1}). Accordingly we have used the modified version of \texttt{CAMB}, discussed in Sec.~\ref{sec:3}. \\

The appearance of temperature anisotropy $\Theta_{\rm ISW}$ on the CMB map, stemming from the evolving gravitational potential $\phi$ over time, can be articulated as follows: 
\begin{equation} \label{eq:20}
\Theta_{\rm ISW} =\frac{\Delta T}{T_{\rm CMB}} \equiv \frac{2}{c^2} \int_{{\rm a}_{\rm dec}}^{1} \frac{\partial\phi}{\partial a} da = -\frac{2}{c^3} \int_{0}^{\chi_{\rm H}} a^2 H(a)\frac{\partial \phi}{\partial a} d\chi
\end{equation}
where $T_{\rm CMB}$ represents the CMB temperature which is equal to 2.725 $K$, $c$ parameterizes the velocity of light, $a_{\rm dec}$ denotes the scale factor of matter-radiation equality, and $\chi$ is the comoving distance defined as:
\begin{equation} \label{eq:21}
\chi (a)=\int_{0}^{1} \frac{c}{a^2 H(a)} da
\end{equation}
Assuming perturbations exist within the horizon, the potential can be related to the matter fluctuation field, using Poisson’s equation in Fourier space, leading to:
\begin{equation} \label{eq:22}
k^2 \phi (k,a)=-\frac{3}{2}a^2 H^2 (a) \Omega_{\rm m} (a) \delta_{\rm m}(k,a),
\end{equation}
where the matter density parameter $\Omega_{\rm m} (a)$ is defined as $\Omega_{\rm m}=\rho_{\rm m}/\rho_{\rm crit}$, where ${\rho_{\rm crit}}$ represents the critical density given by $\rho_{\rm crit}=3 H^2 (a)/8\pi G$ and $\delta_{\rm m} (k,a)$ denotes the matter perturbation.\\
Using Eqs. (\ref{eq:20}) and (\ref{eq:22}) will result in:
\begin{equation} \label{eq:23}
\Theta_{\rm ISW}=\frac{3}{c^3} \int_{0}^{\chi_{\rm H}} a^2 H(a) \frac{\partial \mathcal{F}(a)}{\partial a} \frac{\delta_{\rm m} (k,a=1)}{k^2} d\chi,
\end{equation}
where the $\mathcal{F}(a)$ is defined as:
\begin{equation} \label{eq:24}
\mathcal{F}(a)=a^2 H^2 \Omega_{\rm m} (a) D_+ (a)
\end{equation}
Here, $D_+(a)$ is the linear growth factor which is defined as: $D_+ (a)=\delta_{\rm m} (k,a)/\delta_{\rm m} (k,a=1)$. As long as the dark energy
fluid does not influence local structure formation, the equation governing the evolution of $D_+(a)$ with respect of scale factor is as follows~\cite{linder2003cosmic, peter2009primordial, schafer2008integrated}: 
\begin{equation} \label{eq:26}
\frac{d^2D_+}{d a^2}
+\frac{1}{a}\left(3+\frac{d\ln H}{d\ln a}\right)\frac{d D_+}{d a} = 
\frac{3}{2a^2}\Omega_m(a)D_+(a).
\end{equation}

\begin{figure} [t!]
\begin{minipage}{1\linewidth}
\includegraphics[width=\columnwidth]{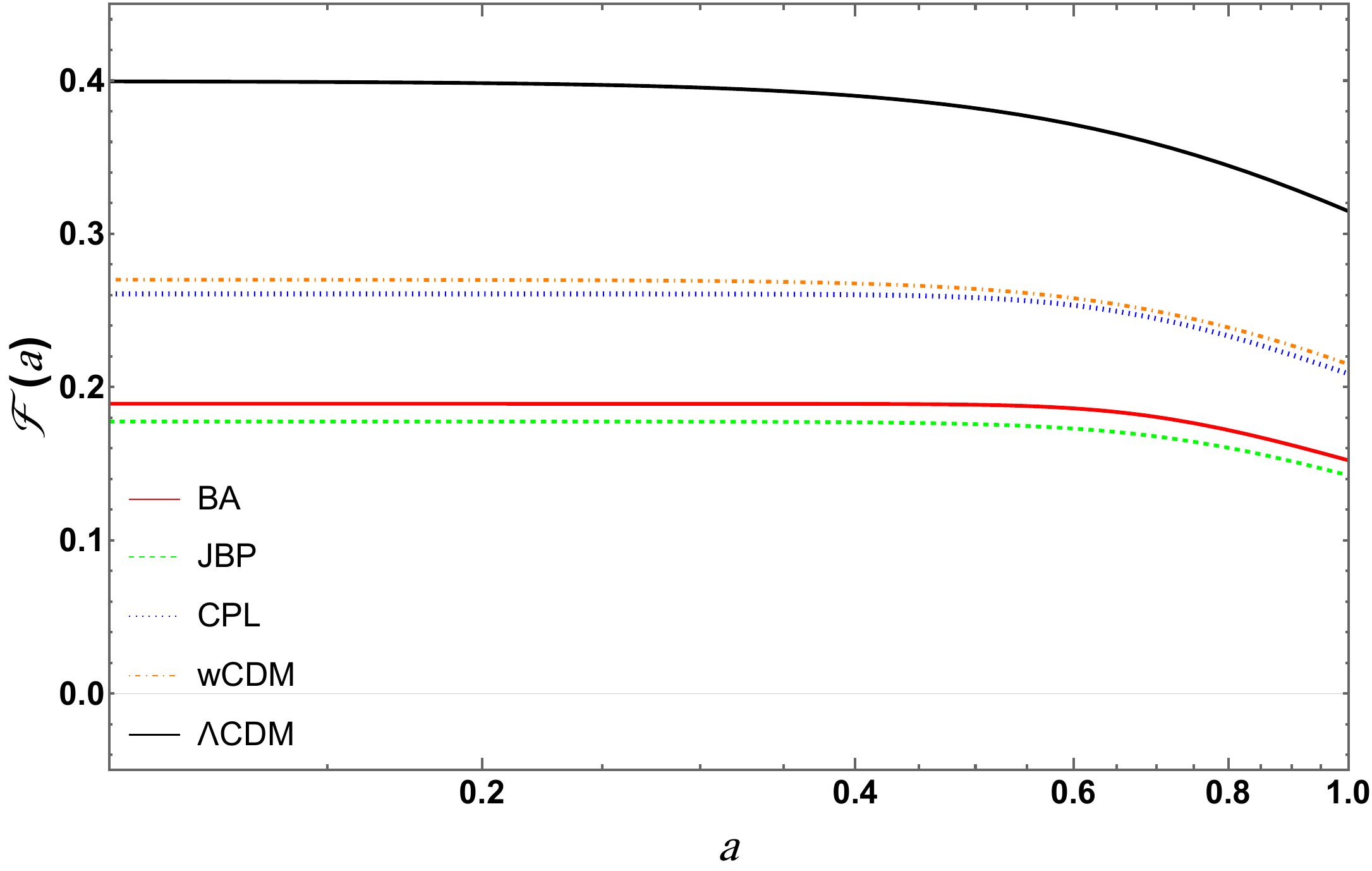} 
\includegraphics[width=\columnwidth]{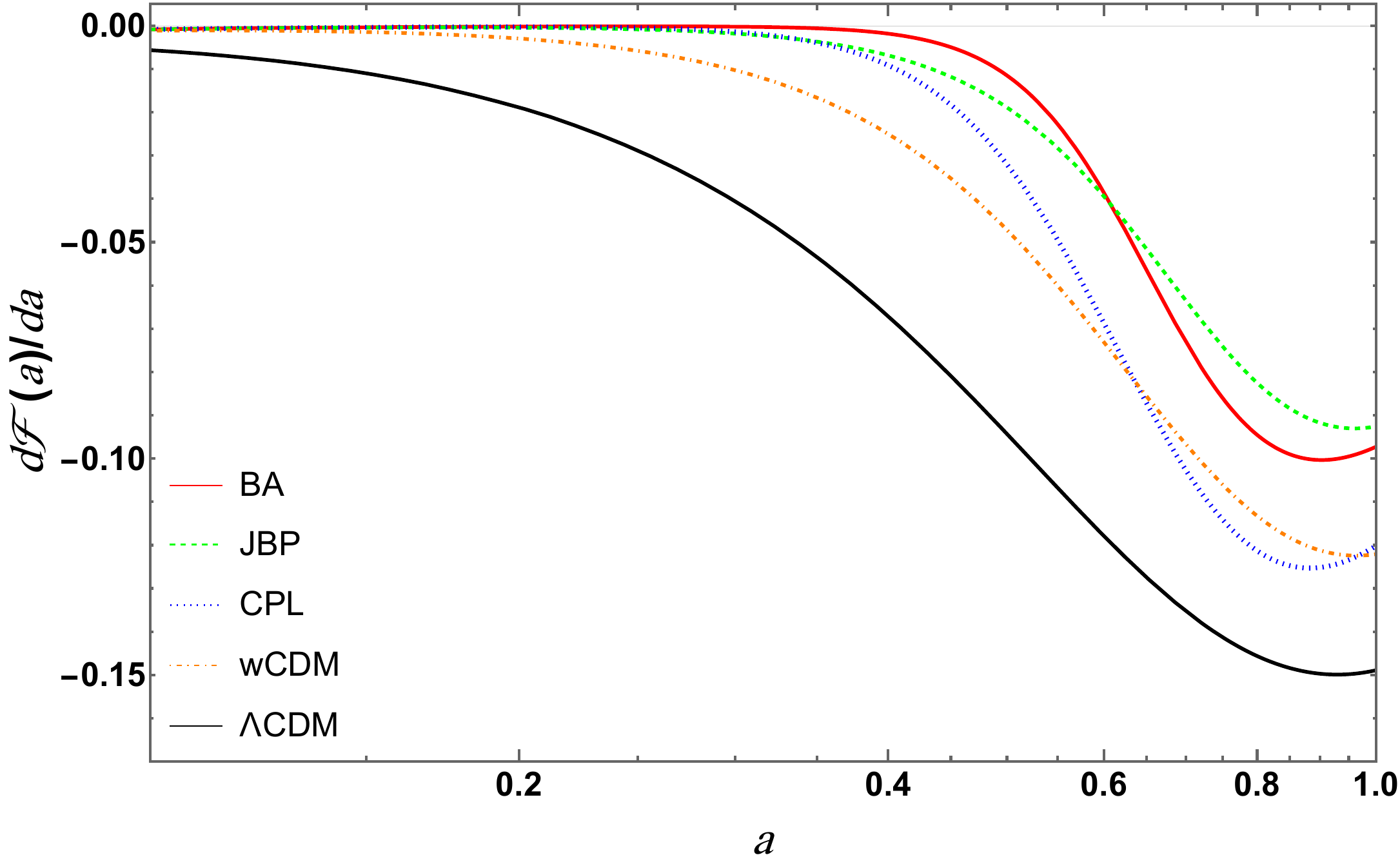} 
\caption{The time evolution of the ISW source term $\mathcal{F}(a)$ (Top) and its derivative $d\mathcal{F}(a)/da$ (Bottom) as functions of scale factor $a$.}
\label{fig:fig4}
\end{minipage}
\end{figure}

Fig.~\ref{fig:fig4} depicts the time evolution of the ISW source term $\mathcal{F}(a)$ (top) and its derivative with respect to the scale factor $d\mathcal{F}(a)/da$ (bottom) as a function of scale factor $a$. The top panel indicates the amplitude of time evolution of the ISW source term as a function of scale factor $a$. Here, it is clear to see the effect of the nature of DE EoS, as the amplitude for those that are dynamical in nature is suppressed in this measurement. Among the constant parameterization of DE EoS, the $\Lambda$CDM model has a higher amplitude compared to the $w$CDM model. As for dynamical DE models, the CPL has the largest amplitude among the dynamical models. As can be seen, the amplitude of the time evolution of the ISW source for the $w$CDM and CPL, as well as the BA and JBP parameterization, are very similar, with the $w$CDM and BA dominating the CPL and JBP, respectively.

However, the impressive feature of the evolution of the ISW source term is that in each model, as we move towards the scales where the DE becomes the dominant force of expansion, a rapid decrease in amplitude can be seen. This derivation with respect to scale factor $a$ is shown in the bottom panel of Fig.~\ref{fig:fig4}. As expected, the most extreme case of the change in time evolution belongs to $\Lambda$CDM. At scale factors, $a \approx 0.6$ and $a \approx 0.65$, a change in the strength of the time evolution of the ISW source term can be observed for the JBP and BA models, as well as the CPL and $w$CDM models, respectively. It should be mentioned that around $a=1$, the values of $d\mathcal{F}/da$ become close to each other once again.

Additionally, it is important to note that in a non-interacting dark sector, the following equation holds for the density parameter :
\begin{equation} \label{eq:27}
\frac{\Omega_{\rm m} (a)}{\Omega_{\rm m}}=\frac{H^2_0}{a^3 H^2(a)} 
\end{equation}
In order to study the potential wells encountered by photons traveling from the last scattering surface to us, it is imperative to evaluate the correlation between the ISW temperature and galaxy density. Acknowledging that, the ISW effect is relatively minor on smaller scales but significantly impactful on larger scales, galaxies serve as a suitable marker for large-scale structures during later cosmic epochs. 

Consequently, the line-of-sight integral for galaxy density is:
\begin{equation} \label{eq:28}
\delta_{\rm g} = \int\frac{H(a)}{c} f(z) D_+(a) \delta_{\rm m}(k,a=1) d\chi 
\end{equation}
The redshift distribution, denoted as $f(z)$, is formulated as $f(z)=b(z)dN/dz$. To understand the vastness of galaxy clusters, the utilization of large-scale surveys becomes necessary in order to compute the abundance of galaxy clusters as a function of redshift $z$.\\
Using the CMB map and galaxy distribution, one can define the angular auto-correlation and cross-correlation as:

\begin{equation} \label{eq:29}
C^{TT} = <\Theta_{\rm ISW} \Theta_{\rm ISW}> 
\end{equation}

\begin{equation} \label{eq:30}
C^{Tg} = <\Theta_{\rm ISW} \delta_{\rm g}> 
\end{equation}
Defining weight functions for photons and galaxies as follows:
\begin{equation} \label{eq:31}
W_{\rm T} (\chi) = \frac{3}{c} a^2 H(a) \frac{d\mathcal{F}(a)}{da}
\end{equation}
\begin{equation} \label{eq:32}
W_{\rm g} (\chi) = \frac{H(a)}{c} f(z) D_+ (a)
\end{equation}
this allows us to write the angular auto-correlation and cross-correlation in a compact notation:
\begin{equation} \label{eq:33}
C^{TT}_{\rm ISW}(\ell) = \int_{0}^{\chi_{\rm H}} \frac{W^2 _{\rm T}(\chi)}{\chi^2}~ \frac{H^4_0}{k^4} ~P (k=\frac{\ell+\frac{1}{2}}{\chi}) ~d\chi
\end{equation}

\begin{equation} \label{eq:34}
C^{Tg} _{\rm ISW}(\ell)= \int_{0}^{\chi_{\rm H}} \frac{W _{\rm T}(\chi) W_{\rm g}(\chi)}{\chi^2} ~\frac{H^4_0}{k^4} ~P (k=\frac{\ell+\frac{1}{2}}{\chi}) ~d\chi
\end{equation}
In this context, P(k) represents the current matter power spectrum and $k=(\ell+1/2)/\chi$ is derived using the Limber approximation for larger value of $\ell$~\cite{afshordi2004cross,stolzner2018updated}

\begin{figure}[t!]
\begin{minipage}{1\linewidth}
\includegraphics[width=\columnwidth]{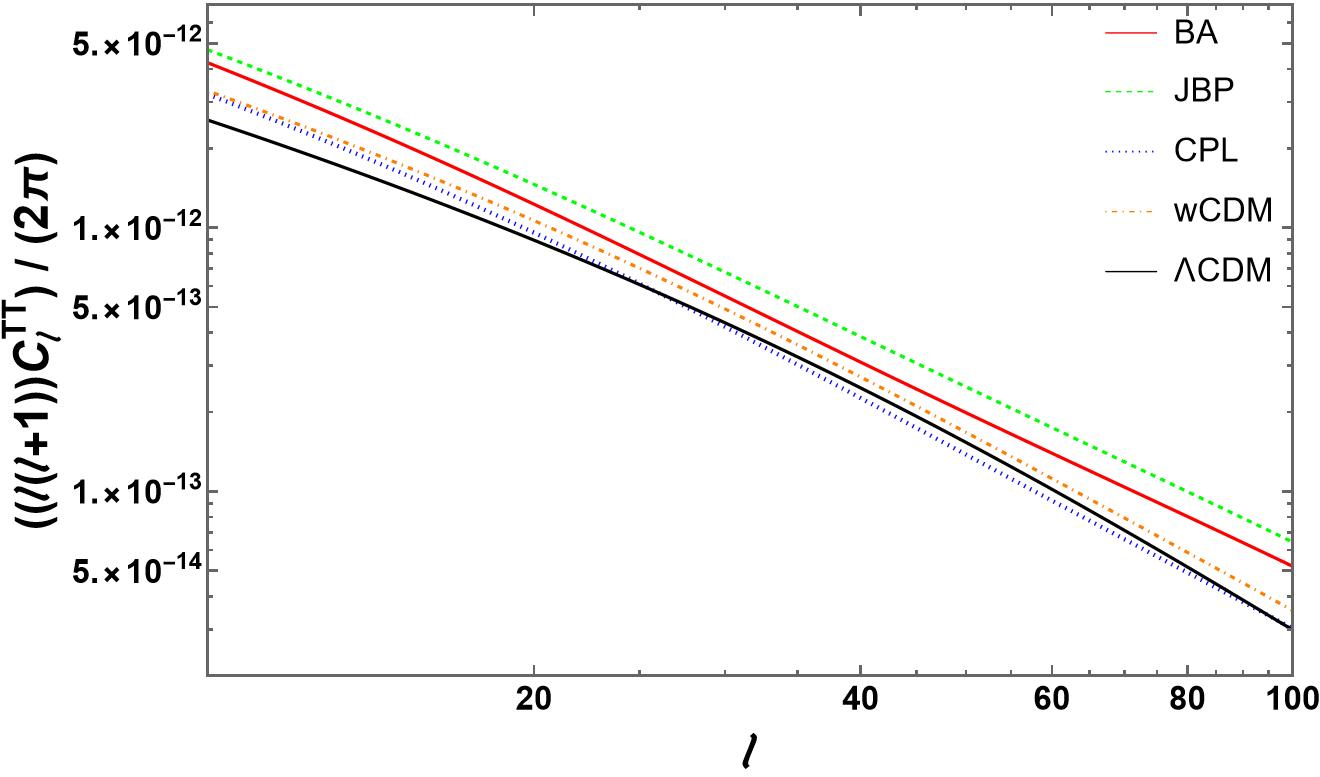}
\caption{The ISW auto-power spectrum $C^{TT}_{\rm ISW}$ as a function of the multipole $\ell$, for BA (red solid line), JBP( green dotted line) and CPL (blue dotted line) as 3 dynamical DE models, as well as $\Lambda$CDM (black solid line) and $w$CDM (dotted-dashed line) as two models of DE with constant equation of state.}
\label{fig:fig5}
\end{minipage}
\end{figure}

\subsection{Redshift Distribution}\label{sec:5.1}
In order to study the correlation between CMB photons and large-scale structure, it is crucial to use information obtained from galaxy redshift distribution surveys. These surveys typically employ spectroscopic and$/$or multi-band photometric calibration survey of small patches of the sky~\cite{sanchez2020propagating}. In this paper, we have used the redshift distribution of four distinct surveys. Hence, in the 
following, we briefly discuss galaxies redshift distribution surveys under investigation. In Sec 4.1 (see Fig. 4) of Ref~\cite{yengejeh2023integrated}, the photometric normalized redshift distribution for the surveys under study are discussed.

{\bf I}. Dark Universe Explorer (DUNE): The main goal of this survey is to investigate potential candidates for weak gravitational lensing and explore the ISW effect. The redshift distribution of the survey is as follows:\\
\begin{equation} \label{eq:35}
f_{\rm DUNE} (z)= b_{\rm eff} [\frac{z_*}{a_*} \Gamma(\frac{3}{a_*})]^{-1} (\frac{z}{z_*})^2 exp [-(\frac{z}{z_*})^{a_*}]
\end{equation}

{\bf II}. The National Radio Astronomy Observatory
(NRAO) Very Large Array (VLA) Sky Survey (NVSS): The redshift distribution function in this survey,
covering about 82\% of the sky, is as follows:
\begin{equation} \label{eq:34}
f_{\rm NVSS} (z)= b_{\rm eff} \frac{a_*^{a_* +1}}{\Gamma (a_*)}\frac{z^{a_*}}{z_*^{a_*+1}} exp (-\frac{a_* z}{z_*})
\end{equation}

{\bf III}. The Sloan Digital Sky Survey (SDSS): This survey gathers images and spectroscopic data of galaxies, quasars, and stars, with the following redshift distribution:
\begin{equation} \label{eq:37}
f_{\rm SDSS} (z)= b_{\rm eff} \frac{a_*}{\Gamma (\frac{m+1}{a_*})}\frac{z^m}{z_*^{m+1}} exp [-(\frac{ z}{z_*})^{a_*}]
\end{equation}

{\bf IV}. Euclid-like: The Euclid mission intends to explore the relationship between the distance and redshift of galaxies up to a redshift of z $\approx$ 2. In this survey, the distribution of redshift is characterized by the following~\cite{weaverdyck2018integrated,martinet2015constraining}:
\begin{equation} \label{eq:38}
f_{\rm Euclid-like} (z)= b_{\rm eff}\frac{3}{2 z^3_*} z^2 exp [-(\frac{ z}{z_*})^{\frac{3}{2}}]
\end{equation}
In the aforementioned surveys, $b_{\rm eff}$, $z_*$, $a_*$ and $m$ are free parameters that need to be determined. $\Gamma(x)$ represents the Gamma function for all surveys. The values for the free parameters are presented in Table~\ref{tab:my_label2}.

\begin{figure*}[t!] 
\centering
\includegraphics[width=0.45\textwidth]{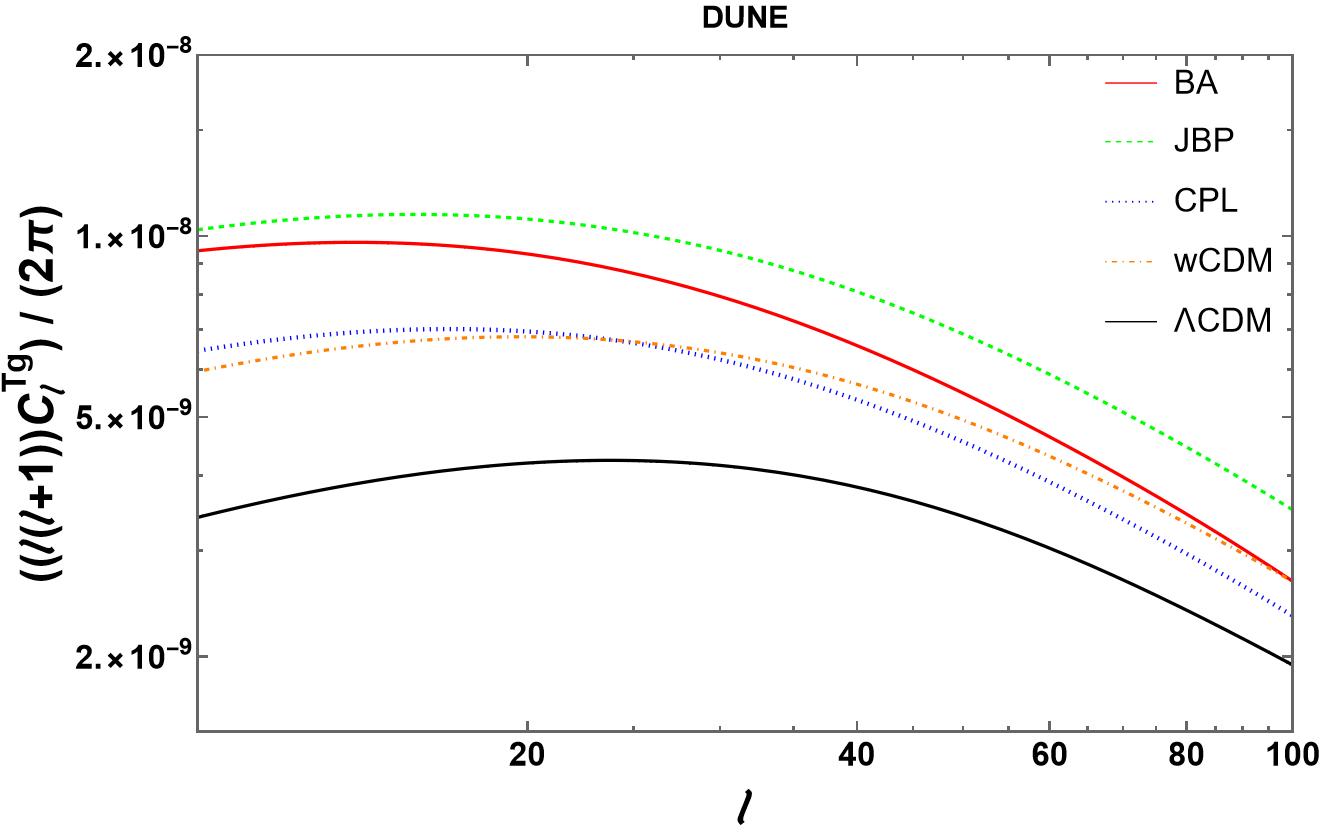}
\includegraphics[width=0.45\textwidth]{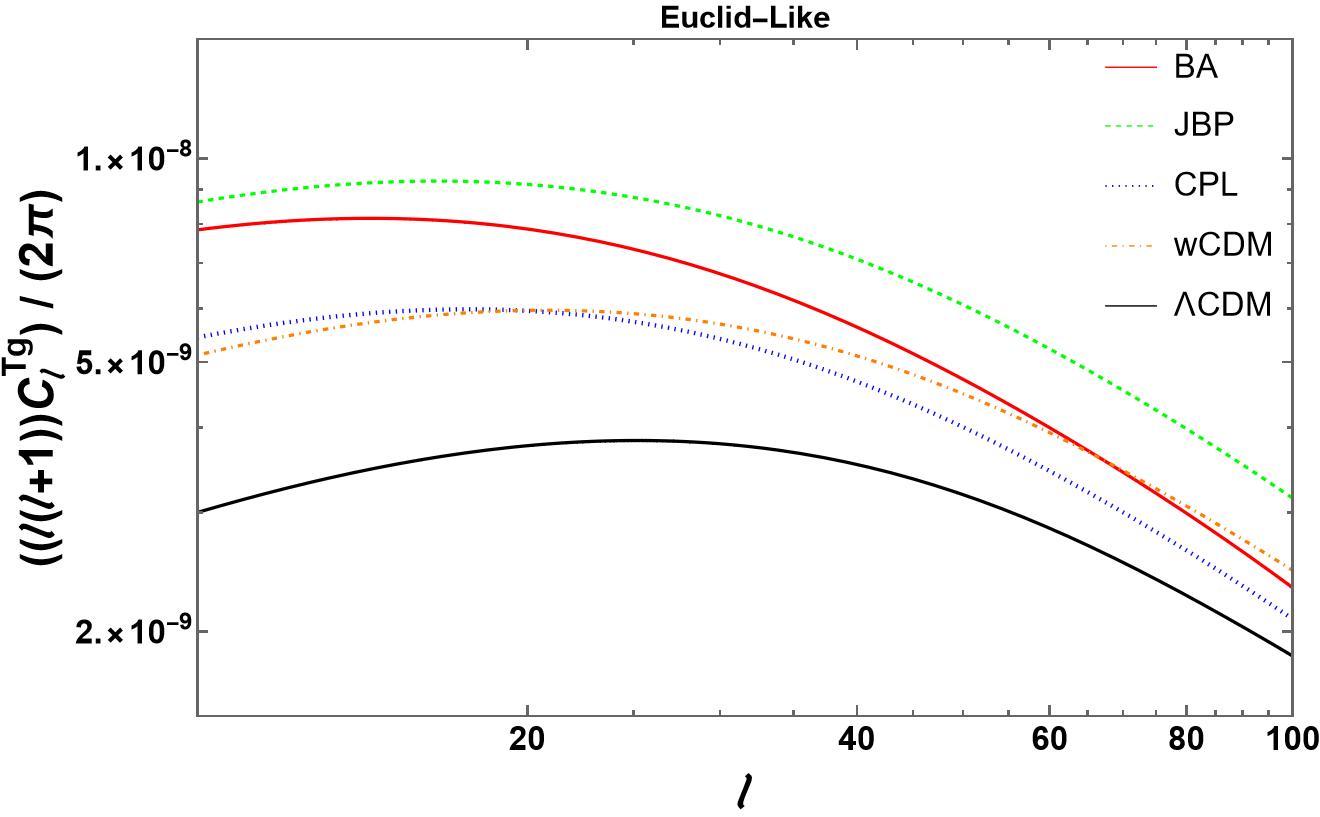}\\~\\
\includegraphics[width=0.45\textwidth]{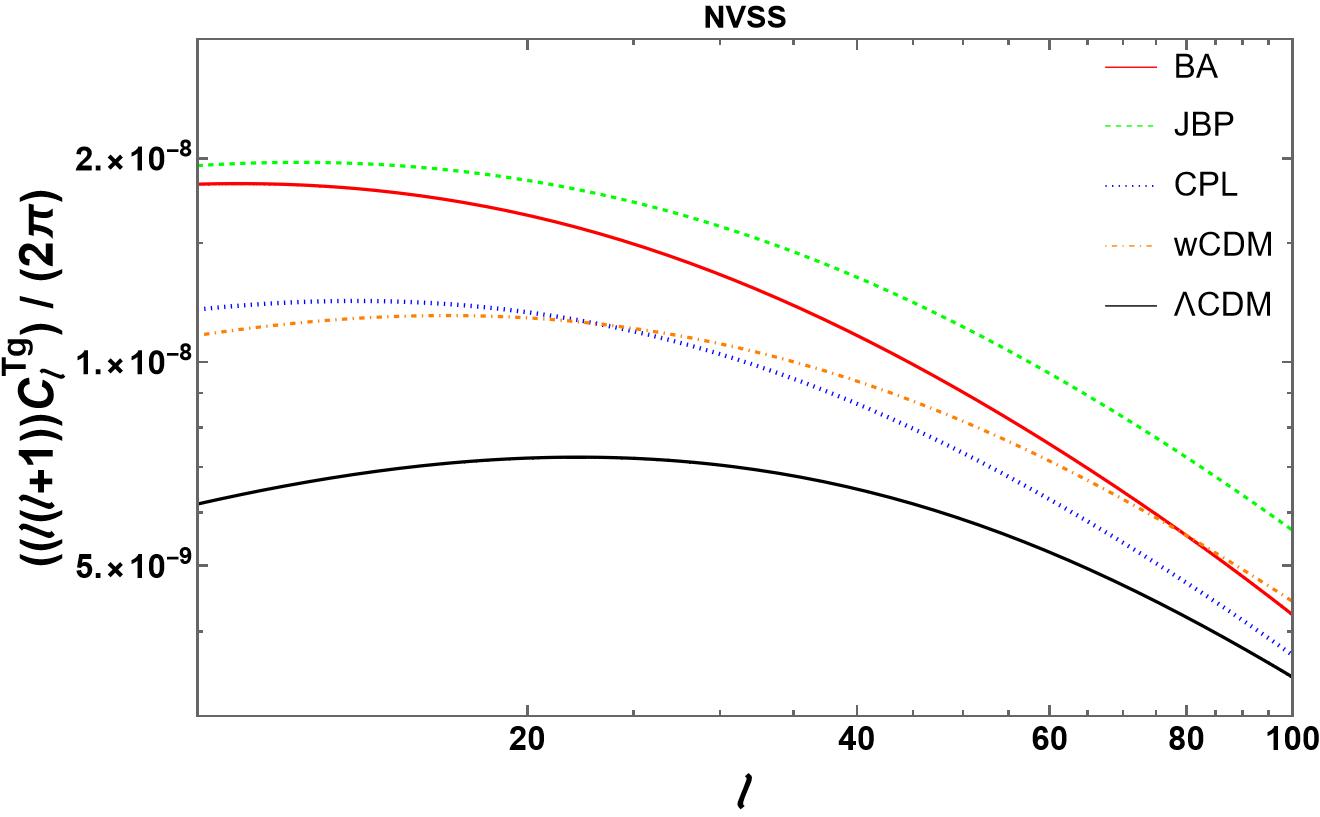}
\includegraphics[width=0.45\textwidth]{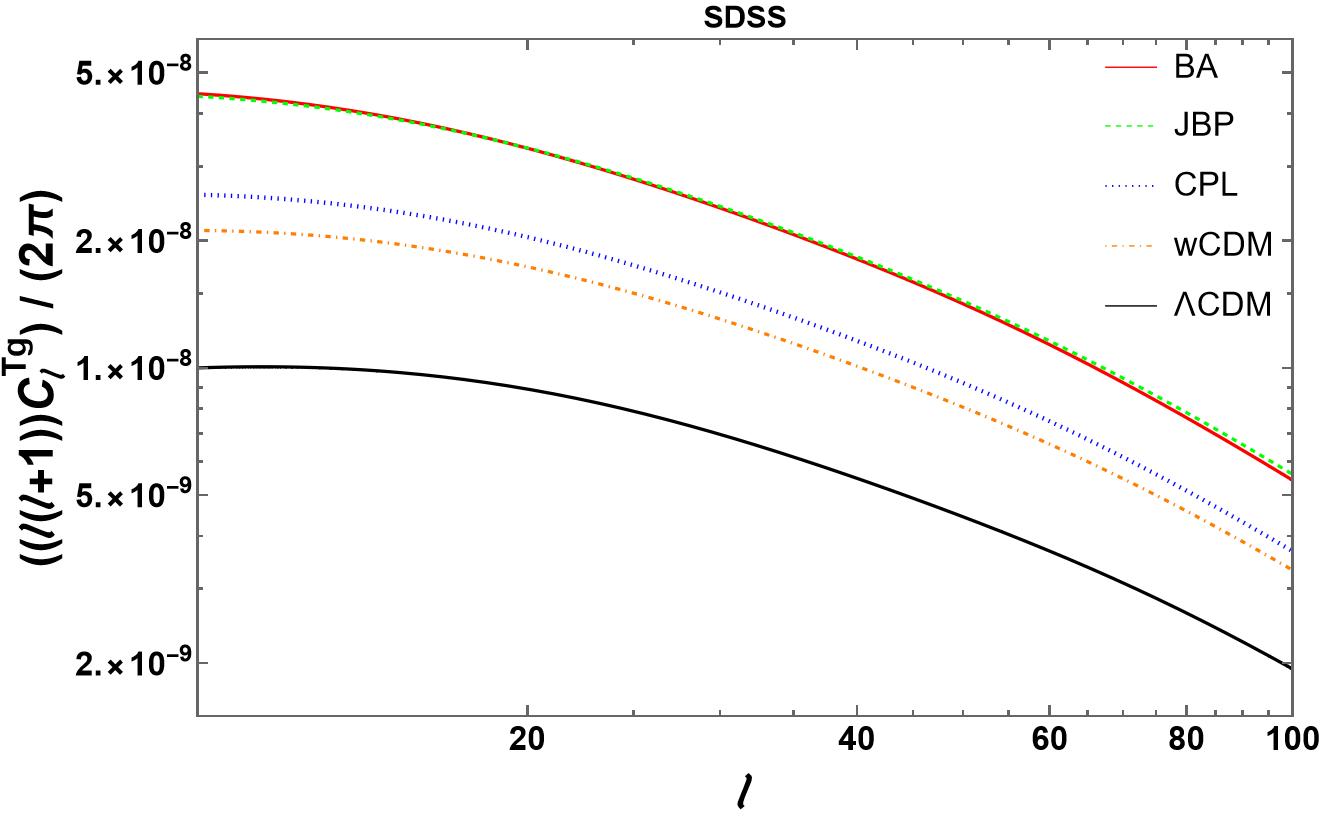}
\caption{The ISW-cross power spectrum $C_{\rm Tg}$ as a function of the multipole $\ell$ in the range of $10$ to $100$, for BA (red solid line), JBP (green dashed line), and CPL (blue dotted line) as 3 dynamical DE models, as well as $\Lambda$CDM (black solid line) and $w$CDM (orange dot-dashed line) as two models of DE with constant EoS, for DUNE (Top left), Euclid-like (Top right), NVSS (Bottom left) and SDSS (Bottom right). }
\label{fig:fig6}
\end{figure*}

\begin{figure*}[t!] 
\centering
\includegraphics[width=0.45\textwidth]{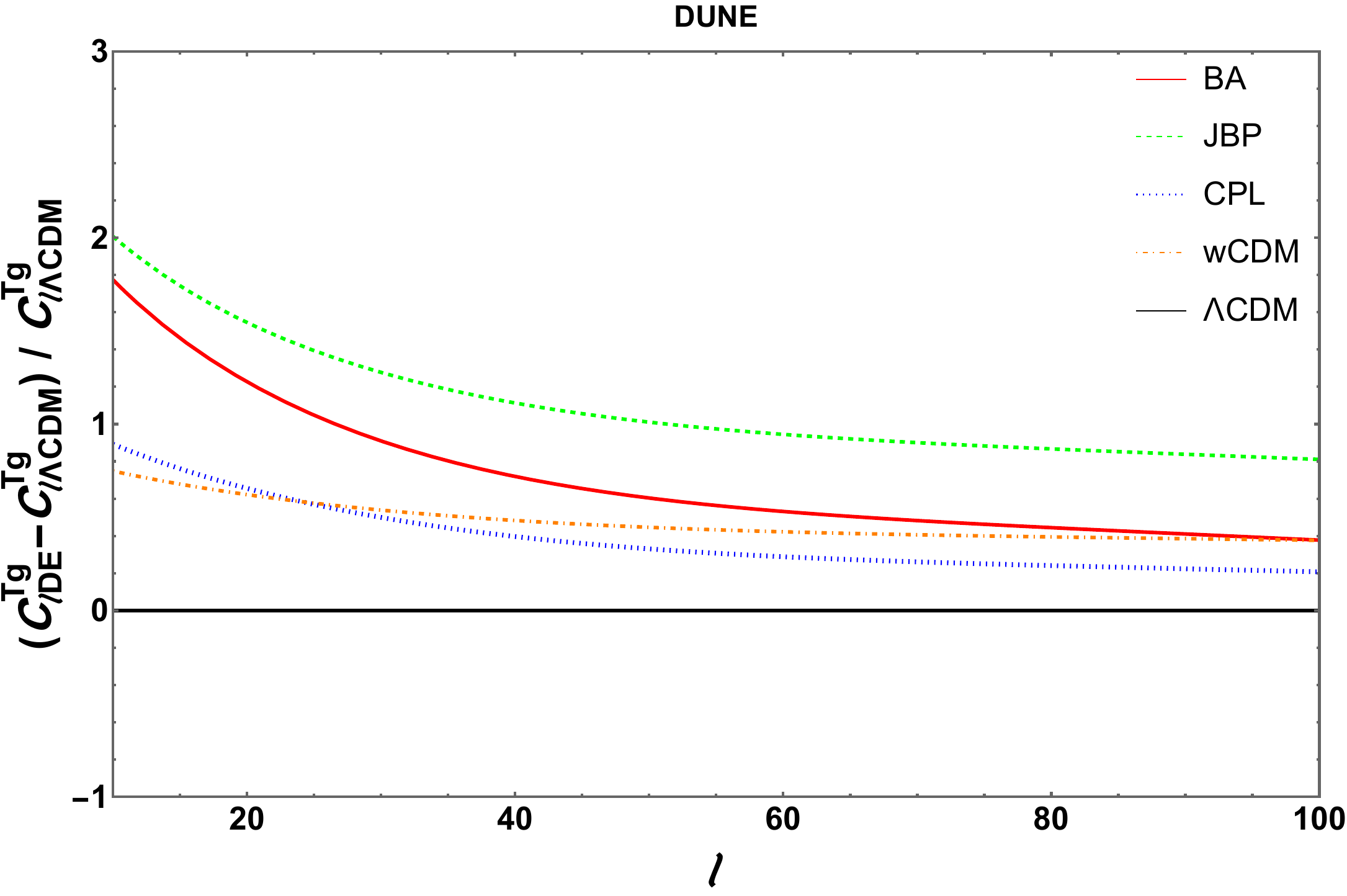}
\includegraphics[width=0.45\textwidth]{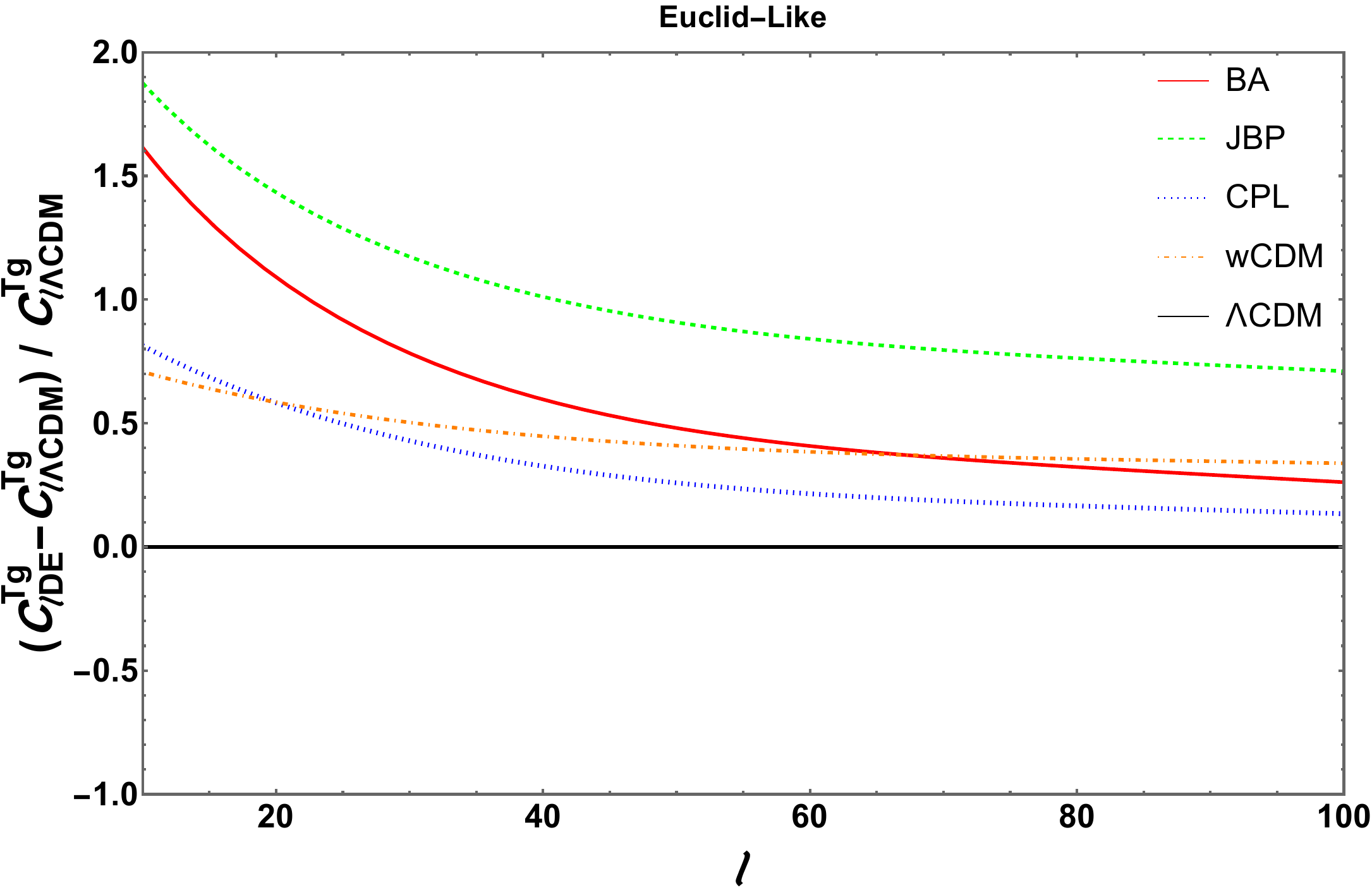}\\~\\
\includegraphics[width=0.45\textwidth]{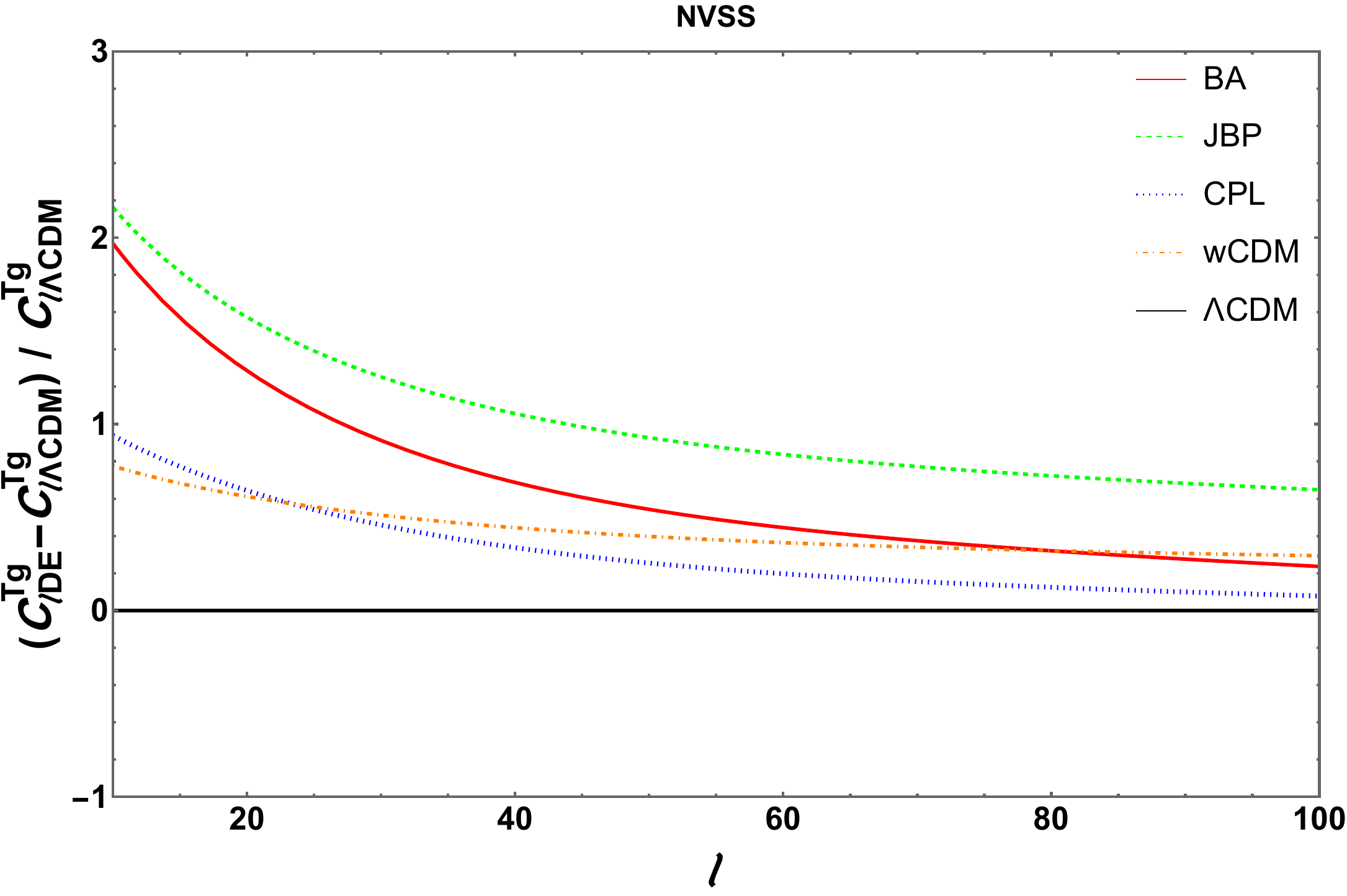}
\includegraphics[width=0.45\textwidth]{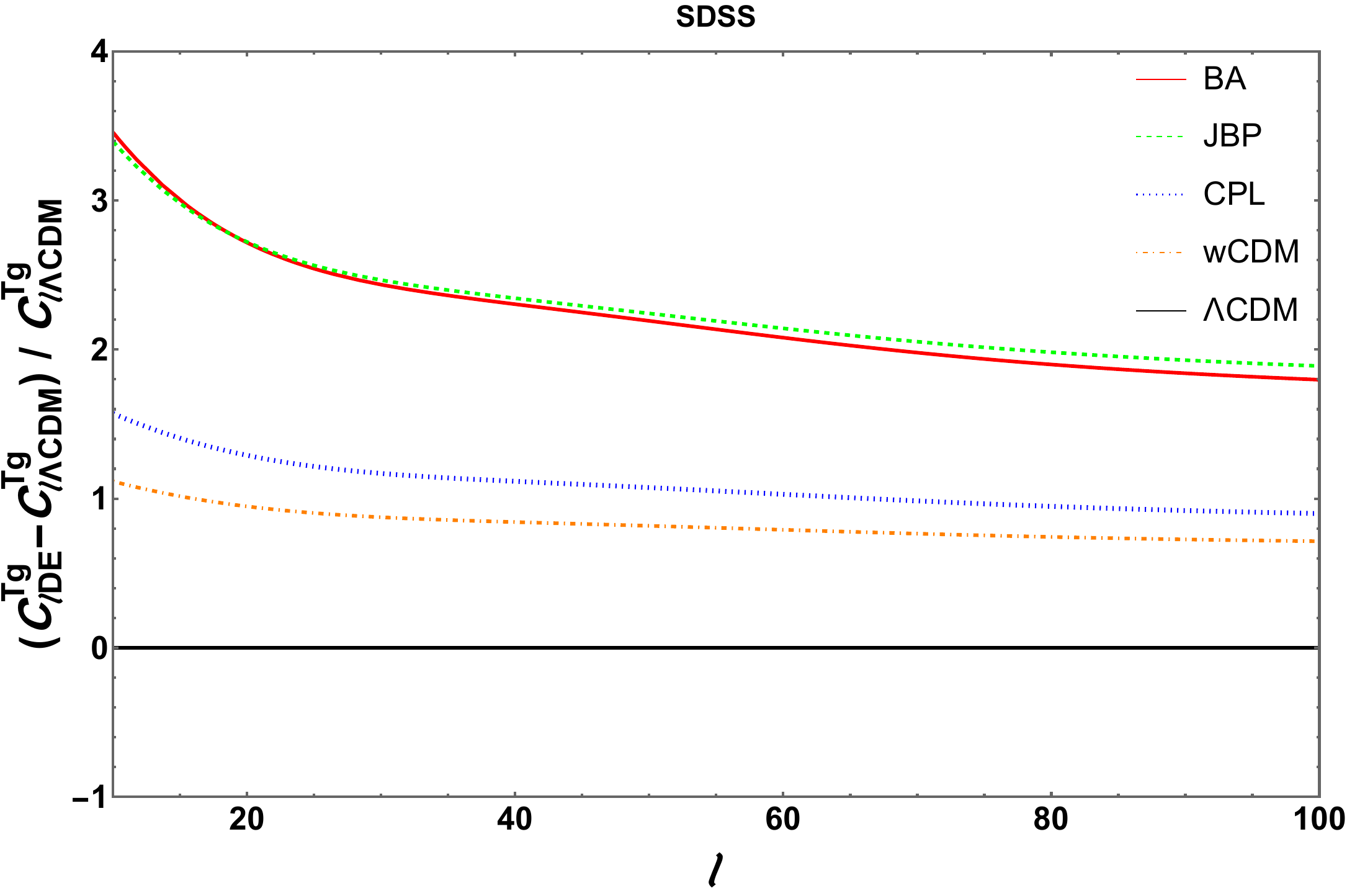}
\caption{Corresponding variation with respect to $\Lambda$CDM for BA (red solid line), JBP (green dashed line), and CPL (blue dotted line) as 3 dynamical DE models, as well as $\Lambda$CDM (black solid line) and $w$CDM (orange dot-dashed line) as two models of DE with constant EoS, for DUNE (Top left), Euclid-like (Top right), NVSS (Bottom left) and SDSS (Bottom right). }
\label{fig:fig7}
\end{figure*}

\begin{table}[!ht]
\centering
\begin{tabular}{c||c|c|c|c}
\hline\hline
\textbf{Survey} & \textbf{$b_{\rm eff}$} & \textbf{$z_*$} & \textbf{$a_*$} & \textbf{$m$} \\
\hline\hline
DUNE & 1.00 & 0.640 & 1.500 & - \\
NVSS & 1.98 & 0.790 & 1.180 & - \\
SDSS & 1.00 & 0.113 & 1.197 & 3.457 \\
Euclid-Like & 1.00 & 0.700 & - & - \\
\hline\hline
\end{tabular}
\caption{ Redshift distribution parameters for DUNE, NVSS, SDSS and Euclid-like surveys.}
\label{tab:my_label2}
\end{table}

\subsection{The ISW auto-power spectrum}\label{sec:5.2}
The ISW auto-power spectrum $C^{TT}_{\rm ISW}$ is shown in Fig.~\ref{fig:fig5} as a function of multipole $\ell(=10-100)$ (inverse angular scale). As depicted, the amplitude of the ISW auto-power spectrum for the JBP and BA models is higher than the other three models, with JBP having the highest amplitude compared to the others. As $\ell$ increases (corresponding to smaller scales), the amplitude of the auto-power spectrum decreases. An intriguing result seen in this figure is that the CPL auto-correlation spectrum amplitude is lower than the $w$CDM model at any given $\ell$. At $\ell \sim 25$ the amplitude of the $\Lambda$CDM auto-correlation spectrum surpasses that of the CPL model, and at $\ell \sim 90$ the amplitudes for $\Lambda$CDM and CPL models start to become indistinguishable from each other. By focusing on the comparison between the CPL and $w$CDM, one can observe that as $\ell$ increases,  the difference in the auto-power spectrum amplitude becomes noticeable, indicating a footprint of the dynamical nature of DE. Moreover, at $\ell \sim 20-80$ we can observe a slight curve in the amplitude of the auto-power spectrum, implying that at these scales, the effects of dynamical DE are most pronounced.\\

\subsection{The ISW-cross power spectrum}\label{sec:5.3}
In Fig.~\ref{fig:fig6}, we have depicted the ISW-cross power spectrum $C^{Tg}_{\rm ISW}$ amplitude as a function of the multipoles $\ell \sim 10-100$. The top left panel represents the ISW-cross power spectrum for the DUNE survey. We can see that the amplitude of the ISW-cross power spectrum amplitude for the JBP models is higher than that of any other models at any multipole $\ell$. The BA model exhibits the next highest amplitude up until  $\ell \sim 90$; it can be observed that as we move toward higher $\ell$ values, the rate at which the BA amplitude weakens is the most pronounced compared to other models. Thus, this effect can be seen in the form of the BA model, with the ISW-cross power spectrum amplitude approaching that of the $w$CDM model. At $\ell \sim 90$, this observed property is best demonstrated as they become indistinguishable at this scale. As depicted in the panel, the CPL model has a higher cross-power spectrum amplitude than $w$CDM at lower $\ell$ values, but as we pass $\ell \sim 20-30$, they converge, and then the $w$CDM model rises in amplitude over the CPL model. It is worth mentioning that the amplitude of $\Lambda$CDM model is lower than that of any other models studied at any $\ell$.\\ 

The top right panel of Fig.~\ref{fig:fig6}, illustrates the ISW-cross power spectrum for Euclid-like survey. The general behavior observed in the DUNE survey holds true for this survey as well, with some exceptions. The first important point is the overall suppression of amplitude compared to the DUNE survey. The second striking feature is the rate at which the BA amplitude is changing with respect to the multipole $\ell$. This feature is so drastic that the BA amplitude crosses below that of the $w$CDM model towards lower amplitudes at $\ell \sim 70$. In this survey, $\Lambda$CDM model also exhibits the weakest ISW-cross power spectrum amplitude.\\

The bottom left panel Fig.~\ref{fig:fig6} shows the NVSS results for the ISW-cross-power spectrum. The change in the amplitude of the ISW-cross power spectrum is an easily observable feature of this panel compared to two other surveys discussed previously. In this panel the prominent features are very similar to those observed in the Euclid-like survey. Here, the crossing point for the CPL and $w$CDM models, as well as the BA and $w$CDM models, is shifted towards higher $\ell$ values ($\ell \sim 25$ and $\ell \sim 80$, respectively). Additionally, the $\Lambda$CDM ISW-cross power spectrum amplitude remains the weakest, similarly to the other two surveys. \\

However, the bottom right panel of Fig.~\ref{fig:fig6}, representing the SDSS survey, shows drastically different characteristics compared to the other three surveys. In this survey, the overall strength of the ISW-cross power spectrum is higher compared to the other three surveys. The strength of the ISW-cross power spectrum for the BA and JBP models is nearly indistinguishable up to $\ell \sim 60$. From that point onward, the JBP model exhibits the highest amplitude compared to other models. Within the compared range of multipoles $\ell$, the CPL model outperforms the $w$CDM and $\Lambda$CDM in the strength of the ISW-cross power spectrum, with $\Lambda$CDM, similar to the other surveys, having the least amplitude among all compared models.\\
Fig.~\ref{fig:fig7} demonstrates the corresponding variation with respect to $\Lambda$CDM model the  for BA, JBP, and CPL as 3 dynamical DE models, as well as $\Lambda$CDM and $w$CDM as two models of DE with constant EoS. This sheds better light on the points discussed earlier.

\section{Conclusion} \label{sec: 6}
In this work, we approached the effect of the nature of DE EoS in two aspects: first, we studied the impact of the DE nature on structure formation via the matter power spectrum at two different redshifts, $z=0$ and $z=5$. Second, we observed the ISW effect on the scale of $\ell = 10-100$ by analyzing the ISW auto-correlation power spectrum $C^{TT}$ and the ISW cross-correlation power spectrum $C^{Tg}$. 

In the first section of this study, with the best-fit values used in this work, one can observe that regardless of the model, as the redshift increases, the amplitude of the matter power spectrum is suppressed, confirming that the strength of structure formation is highest in each model at the present day. Moreover, from Fig.~\ref{fig:fig2} and~\ref{fig:fig3}, one can infer that the strength of the matter power spectrum for models with a dynamical nature in their EoS is stronger compared to those with a constant nature. Among those with a dynamical EoS, the JBP model exhibits the highest amplitude of the matter power spectrum, while the CPL model has the weakest; to the extent that it is indistinguishable from $w$CDM model (Bottom panel of Fig.~\ref{fig:fig2}).

In the second part of this work, we focused on the ISW effect within the framework of dynamical DE in comparison with models featuring constant DE equations of state. First, in Fig.~\ref{fig:fig4} we have shown the time evolution of the ISW source term $\mathcal{F}(a)$ and its derivative $d\mathcal{F}(a)/da$ as functions of the scale factor $a$, which reveals a stronger amplitude of $\mathcal{F}(a)$ for models with constant DE EoS. Here, $\Lambda$CDM exhibits the highest amplitude among the models with constant DE equations of state overall, while CPL exhibits the highest amplitude among the dynamical models, whereas JBP has the lowest. Additionally, as we move towards the scales where DE becomes the dominant force, one can observe a decrease in the amplitude of $\mathcal{F}(a)$. This effect can be seen in the right panel of Fig.~\ref{fig:fig4}, which depicts the time evolution of the ISW source term with respect to the scale factor $a$. Fig.~\ref{fig:fig4} also indicates that in all cases studied, DE becomes the dominant force in the late-time universe, but the scale at which the effect of DE on $\mathcal{F}(a)$ becomes noticeable is highly dependent on the characterization of the DE equation of state. Together, Fig.~\ref{fig:fig4} in combination with Fig.~\ref{fig:fig2} and~\ref{fig:fig3}, can help one observe the dynamical effect of DE on the essence of structure formation.

It is known that the ISW effect can be represented by the auto-correlation power spectrum and the cross-correlation power spectrum between 
the CMB temperature and large-scale structures such as
galaxies. However, the contributions from auto-correlation are subdominant compared to the primordial contributions, hence they cannot be detected by observational data. On the other hand, the cross-correlation is large enough to be detected in various studies~\cite{enander2015integrated}.

In this study, Fig.~\ref{fig:fig5} helps to observe the strength of the ISW auto-power spectrum, where the JBP and BA models rank the highest (with JBP being the highest). One interesting outcome was that in this dataset of choice, the $w$CDM model favors higher amplitude over the CPL parameterization in the range of $\ell$s under study. In Fig.~\ref{fig:fig5}, one can even observe that after passing $\ell \sim 25-30$, $\Lambda$CDM surpasses the CPL model in amplitude.

At the end, we have analyzed the ISW-cross power spectrum for 4 different surveys (DUNE, Euclid-like, NVSS and SDSS). The three panels of Fig.~\ref{fig:fig6} that show DUNE, Euclid-like, and NVSS surveys indicate that the highest values for the ISW-cross power spectrum belong to the dynamical DE models, specifically the JBP and BA models. However, one can expect a shift between the dominance of the CPL and $w$CDM models. Using the SDSS redshift distribution will reveal a different characteristic. First the amplitude of the ISW-cross power spectrum is higher compared to other surveys. Second, the dynamical DE models always dominates over DE model with a constant DE equation of state. Thus, it is difficult to discriminate between the BA and JBP parameterizations, especially at lower $\ell$s. It should be noted that in all surveys, the $\Lambda$CDM model is subdominant in terms of the strength of the ISW-cross power spectrum. This study has shown that both the matter power spectrum and the ISW signal provide valuable information for distinguishing the nature of DE. Fig.~\ref{fig:fig7} illustrates the corresponding variation with respect to $\Lambda$CDM, confirming the earlier discussion.

It is worth mentioning that in this study, we have focused on the linear approximation of the matter power spectrum. However, studying the non-linear matter power spectrum could provide key information regarding the dynamical nature of dark energy, especially on smaller scales (larger $k$). Additionally, it would be informative to study the matter power spectrum at higher redshifts by incorporating datasets that contain information from redshifts deep into the matter-dominated era. An intriguing idea that could spark new discussions is to investigate the matter power spectrum in the future, as at that point, the dynamical nature of dark energy could leave its footprint more prominently over time.

\section*{ACKNOWLEDGMENTS}
MR and MN gratefully thank Saeed Fakhry and Mina Ghodsi  for the useful discussion and constructive comments.\\
EDV acknowledges support from the Royal Society through a Royal Society Dorothy Hodgkin Research Fellowship. This publication is based upon work from COST Action CA21136 – ``Addressing observational tensions in cosmology with systematics and fundamental physics (CosmoVerse)'', supported by COST (European Cooperation in Science and Technology).

\bigskip

\bibliography{Folder}

\end{document}